\documentclass[aps,reprint]{revtex4-1}

\usepackage{graphicx}

\begin{document}

\title[]{Common surface structures of graphene and Au(111) : \\
The effect of rotational angle on adsorption and electronic 
properties}

\author{Merve Yortanl{\i}}
\author{Ersen Mete}\email[Corresponding author : ]{emete@balikesir.edu.tr}
\affiliation{Department of Physics, Bal{\i}kesir University, Bal{\i}kesir 10145, 
Turkey}
\date{\today}
\begin{abstract}
Graphene adsorption on Au(111) surface was explored to identify
their common surface structures by means of van der Waals corrected density 
functional theory calculations. The alignment of graphene in the form of 
certain rotational angles on the gold surface has an important role on the 
lattice matching which causes Moir\'{e} patterns, and on the electronic 
properties of the resulting common cell structures. The dispersive weak 
interactions between carbon and gold layers lead to a downward shift of Fermi 
energy of the adsorption system with respect to the Dirac point of graphene 
showing a $p$-type doping character. Moreover, the shift was shown to depend on 
the rotational angle of graphene on Au(111).
\end{abstract}


\maketitle

\section{Introduction}
Graphene is a two dimensional material made up of carbon atoms in a hexagonal 
lattice arrangement. It receives a growing attention since its discovery in 
2004 \cite{Novoselov1,Novoselov2}, due to its unusual mechanical, thermal, 
electronic and optical properties (e.g. impressive mechanical strength and 
flexibility, high thermal conductivity, ultrahigh charge-carrier mobility, 
and unique optical response).\cite{Nair,Sreekanth,Bolotin} 

Theoretically, graphene can be modeled as a free-standing two dimensional 
sheet. In this form, graphene's superior electronic conductivity is related 
to the conical singularities, at the corners of the hexagonal Brillouin zone, 
which show linear energy dispersion up to $\pm$1 eV from the Fermi level.
These structures also known as the Dirac cones where the upper and lower 
conicals from the conduction and valence bands touch at the Dirac point.
This gives graphene a gapless electronic structure and massless charge-carriers 
which lead to novel topological phenomena such as quantum spin Hall 
effect.\cite{Kane,Neto}  

The growth and transfer of free-standing graphene onto metal surfaces in 
control-based electronic device designs brings some difficulties about 
scalable production, desirable quality, and practicality. Moreover, 
design specific functionality to graphene can be achieved either by dopants 
(such as boron nitride (BN) molecules, H, Al, Si, P and S atoms), or by 
adsorbing the graphene monolayer onto a metal surface, or even by using 
graphene nanostructures like 
nanoribbons.\cite{Gweon,Ci,Denis,Nilsson,Wintterlin,Han,Chen}  

The increasing demand for functional electronic applications, such as display 
technologies, pushes researchers to develop substrate supported transparent 
conducting layers. Graphene monolayers with long-range order can be synthesized 
on metal surfaces via, for example, chemical vapor deposition 
(CVD).\cite{MHKang} Graphene coating on metal surfaces is an advantageous 
method which can be adapted to a wide range of areas such as electronics, 
solar cells, optoelectronics, sensor-technology and bio-devices. \cite{THHan, 
Oostinga, XLi, Reina, Nouchi,Salihoglu,KKim,SLin} 

The growth of graphene monolayers on metal foils such as Cu 
leads to polycrystalline films with domains depending on the 
crystallographic orientations. Moir\'{e} superstructures were observed 
on epitaxial graphene which occur due to matching of graphene lattice with that 
of the metal substrate depending on the periodic coincidence and the rotational 
angle between them.\cite{Wintterlin,Merino} Wofford \textit{et 
al.}\cite{Wofford} synthesized and observed graphene orientation in a R30 
alignment on Au(111) using low-energy electrron diffraction (LEED). In this 
study, one of the aims is to identify a number of small size common surface cell 
structures of graphene and gold (111) by considering possible rotational 
alignments. The choice of gold as the substrate has some advantages such as 
its low C solubility, inert surfaces to oxidation, lower vapor pressures 
relative to that of Cu, and re-usability.\cite{Wofford}

The adsorption and electronic properties of graphene/metal surface structures 
must be well understood for highly efficient device designs. For this 
reason there are many experimental studies to elaborate the mechanism of 
graphene adsorption on metal surfaces. \cite{Pletikosic, Kwon, Vita, Dedkov, 
Marchini, Karpan, Gao,Enderlein, Moritz, Varykhalov, Peter, Peter1, Klusek, 
Prezzi} For instance, the interaction between graphene and Ti(0001), Ni, Pd 
and Co (111) surfaces are known to be stronger (chemisorption) relative to 
physisorption on Au, Ag, Pt, Al, Ru surfaces. The reasons leading to these 
two adsorption regimes, and their affects on the chemical and electronic 
properties of grapheme/metal systems have been investigated by many theoretical 
studies.\cite{Giovannetti1, Slawinska1, Slawinska2, Bertoni, 
Hamada, Gong2010, Khomyakov, Kang, Brako, Kozlov, Swart, Uchoa, Fuentes, Ugeda, 
Xu, Vanin, WZhao, Andersen}  The Dirac conical structure of graphene is 
less affected at the weak interfacial interaction regime. However, due to the 
difference in graphene and metal surface work functions, Fermi level of the 
total system shifts up relative to the Dirac point on Ag, Al, Cu substrates 
($n$-type doping) or down on Au, Pt substrates ($p$-type doping). The upward or 
downward shift of Fermi level is considered to be associated with the work 
functions of graphene and the corresponding transition metal 
surface.\cite{Khomyakov,JZheng} In addition, the equilibrium distance 
between adsorbed graphene and the metal surface determines the Fermi level 
shift.\cite{Slawinska1} Chemisorption on Ni, Pd and Co surfaces, on the other 
hand, is due to the strong interaction between the $p_{z}$ orbitals of C atoms 
graphene and the $d$ orbitals of the metal surface. In this case, the band 
structure of graphene changes and mixes with that of metal 
substrate.\cite{Giovannetti1, Khomyakov} 

The computational studies usually consider (2$\times$2) supercell of graphene as the 
smallest common  structure for graphene and metal surfaces. There is a need to 
determine the possible common supercell structures with different periodicities. 
In this study, the metal substrate has been chosen as gold. Apart from the 
advantageous sides of gold as a substrate for the adsorption of graphene 
monolayers mentioned earlier, Au(111) is a well-known surface which is encountered 
in many applications such as self-assembled monolayers (SAM).\cite{Dimilla, 
Kumar, Gorman, Mrksich, Yamada, Schon, Chaki, Lao} In recent years, there are a 
few studies on the application of graphene coating on gold surface-based 
SAMs.\cite{Xie, Li2018, Yan} For instance, Xie \textit{et al.} comparatively 
analyzed the electrochemical properties of bare gold, of gold with n-octadecyl 
mercaptan (C$_{18}$H$_{37}$SH) SAMs, and of graphene adsorbed SAM/Au 
system.\cite{Xie} They reported that there was no electron transfer between the 
non-graphene/SAM structure and the ruthenium hexamine redox probe, so the 
material exhibited insulating behavior. However, they observed an electron 
transfer between the graphene coated SAM and the redox probe. In another recent 
study, Yan \textit{et al.} studied the photoemission characteristics of gold 
surface/diamondoid SAMs covered with graphene as the stabilizer/protector 
layer.\cite{Yan} They reported that the graphene coating prevents the desorption 
of diamondoid molecules from the gold surface by forming a good barrier over the 
diamondoid SAMs. Such applications put forward the need to determine 
common surface supercells of hetero-layered structures which involve Au and graphene. 
The combination of gold and graphene has a great potential in different types 
of novel designs.

We systematically investigated the matching periodicities of graphene on Au(111) 
to determine their several common surface cells using modern dispersion 
corrected DFT calculations. One of the main focuses has been given to elaboration 
of the adsorption geometries, interlayer binding characteristics, energy bands, and 
work functions of these structures. Since different common geometries are expected 
to show various atomic coincidences, it becomes important to explore the effect of 
such differences between the common periodicities on the adsorption and electronic 
properties of graphene on Au(111). 

\section{Theoretical Method}

Total energy density functional theory (DFT) calculations in the framework of 
the projected augmented wave method (PAW) have been performed using VASP 
\cite{Kresse,Furth,Joubert,Blochl}. Single-particle orbitals were expanded in 
the plane wave basis up to a kinetic energy cut-off value of 400 eV. The 
exchange-correlation (XC) effects have been included based on the modern and 
nonempirical SCAN\cite{SCAN} (strongly constrained and appropriately 
normed) semilocal meta-Generalized Gradient Approximation (meta-GGA) density 
functional. 

To achieve more consistent results for weakly interacting layered structures, 
van der Waals (vdW) forces must be taken into consideration. As known from the 
previous studies, the interaction between Au surface and graphene layer is a 
weak physisorption interaction.\cite{Giovannetti1,Nie} Hence, we have 
used SCAN+rVV10 exchange-correlation functional which includes dispersive 
corrections through non-local correlation functional rVV10.\cite{rVV10} In fact, 
recent studies showed that the SCAN+rVV10 vdW density functional produces 
excellent results for interlayer binding energies of layered materials as well as the
adsorption energies of graphene on transition metal surfaces.\cite{Peng}
In order to give a better description of the effect of the vdW-DFT approach on the
computationally estimated values of the weakly interacting graphene/gold system, 
the results obtained by using PBE functional, which is based on the standard 
generalized gradient approximation (GGA), are also presented.

In addition, even though gold is a metallic system, use of different vdW 
functionals make a difference in the calculated lattice constant. As given in 
Table~\ref{table1}, the SCAN+rVV10 functional yields the best estimation to 
the experimental value of 4.078 {\AA}. \cite{Wyckhoff} The in-plane lattice 
constant of graphene was found as 2.445 {\AA} and 2.442 {\AA} with PBE 
and SCAN+rVV10 functionals in agreement with previous studies.\cite{Khomyakov}

\begin{figure}[h]
\includegraphics[width=6cm]{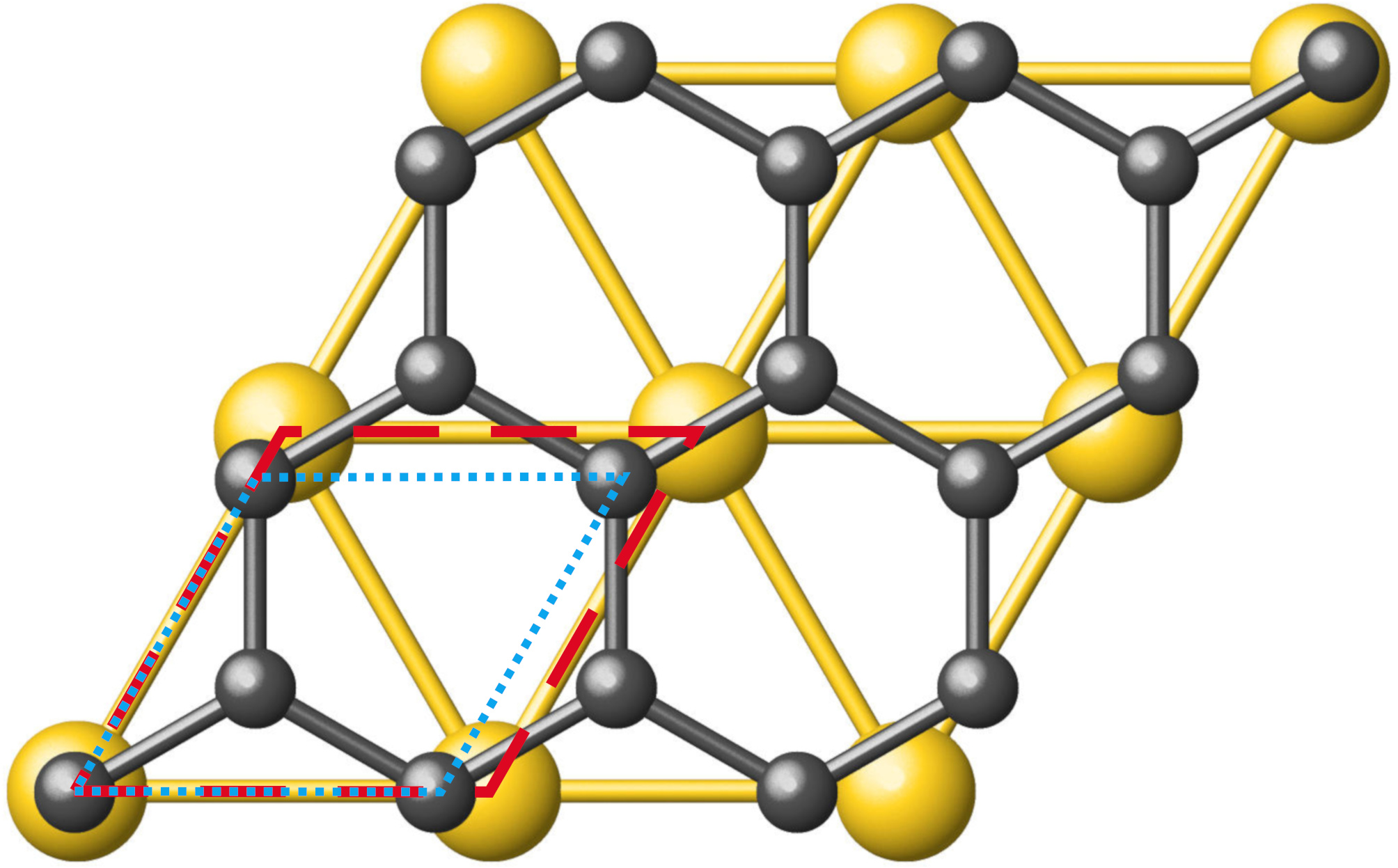}
\caption{\label{fig1} Schematic representation of the graphene layer on the top 
layer of Au(111) surface. Dotted (blue) and dashed (red) lines indicate the 
unitcells of graphene and Au(111), repectively.}
\end{figure}

\begin{table}[h]
\caption{\label{table1}Comparison of the bulk Au lattice constant between 
experimental and various XC functionals values.}
\begin{ruledtabular}
\begin{tabular}{@{\hspace{5mm}}lc@{\hspace{5mm}}}
Method & Au lattice constant (\AA)\\
\hline
Exp.\cite{Wyckhoff}  &4.078 \\
LDA          &4.052 \\
PBE          &4.156 \\
PBE-DF2      &4.330 \\
PBE+dDsC     &4.110 \\
SCAN         &4.095 \\
SCAN+rVV10   &4.073 \\
\end{tabular}
\end{ruledtabular}
\end{table}

A four layer slab model was built to represent (111) surface of gold. Graphene is 
placed on gold  surface such that their lattice translation vectors coincide as 
shown in Fig.~\ref{fig1}. The magnitude of the in-plane translation vectors of gold 
and graphene were found by employing vdW corrections as 2.744 {\AA} and 
2.442 {\AA} , respectively. Therefore, (1$\times$1) unit cell of graphene has a 
11\% mismatch to that of gold. By rotating graphene monolayer on gold surface, 
matching supercell structures were explored. A strain of about \%2 on graphene 
caused by adsorption on the gold surface is imposed as a criteria to identify 
common supercells. The strain percentage on graphene is calculated using,
\[
\frac{d_{\mathrm{C-C}}^{\,\mathrm{free}}-d_{\mathrm{C-C}}^{\,\mathrm{adsorbed}}}
{d_{\mathrm{C-C}}^{\,\mathrm{free}}}\times100
\] 
where $d_{\mathrm{C-C}}^{\,\mathrm{free}}$ and $d_{\mathrm{C-C}}^{\,\mathrm{adsorbed}}$ 
are the C-C bond lengths in free-standing and adsorbed graphene monolayers.

Each of the computational cells consists of a four layer gold slab with a graphene 
adsorption monolayer obeying the same periodicity, and  a 12{\AA} thick vacuum 
region to prevent any unphysical interaction between the periodic images of the slabs. 
Methfessel-Paxton smearing with $\sigma$= 0.05 was used in the calculations. Atomic 
coordinates were optimized self-consistently until Hellman-Feynmann forces 
acting on each atom at each of the three cartesian directions became less than 
10$^{-2}$ eV/{\AA}. No atom was frozen to its bulk position. In the geometry 
optimizations, surface Brillouin zone (BZ) integrals were carried 
out by $k$-point samplings which were chosen properly to be dense enough for 
a metallic system and to be compatible with the translational symmetry of the 
corresponding reciprocal cell. In the structure optimizations we switched off the 
symmetry operations which reduces the number of $k$-points in the irreducible BZ. 
For example, $\Gamma$-centered $6\times6\times1$  $k$-point grid was used for 
the (4$\times$4)  computational cell which has a hexagonal symmetry. The naming 
of the common supercell structures is hereinafter adapted with respect to the (111) 
surface of  gold. Similarly, an $8\times8\times1$ $k$-point mesh was chosen 
for ($3\times3$)  and  ($\sqrt{7}\times\!\sqrt{7}$)R19.12$^\circ$ supercells.  
For the ($\sqrt{3}\times\!\sqrt{3}$)R30$^\circ$ structure $10\times10\times1$ $k$-point 
sampling was enough to get energy values converged to a few meVs. In the relaxation 
of the ($7\times 7$) superstructure, the BZ integrations were performed over a 
$4\times4\times1$ $k$-point grid. 

In order to describe the physisorption of a graphene monolayer on Au(111) with 
different possible common periodicities, the average adsorption energy 
E$_{\mathrm{ads}}$ per carbon can be obtained using the following formula,
\[
E_{\mathrm{ads}}=(E_{G/Au(111)}-E_{Au(111)}-E_{G})/n
\]
\noindent where E$_{G/Au(111)}$ is the total cell energy of graphene physisorbed
Au(111) slab. E$_{Au(111)}$ and E$_{G}$ are the energies of
clean gold surface model and the graphene monolayer, respectively. In addition 
$n$ in the denominator is the number of carbon atoms contained in the 
computational cell. 

\begin{figure*}[htb]
\centering
\includegraphics[width=15cm]{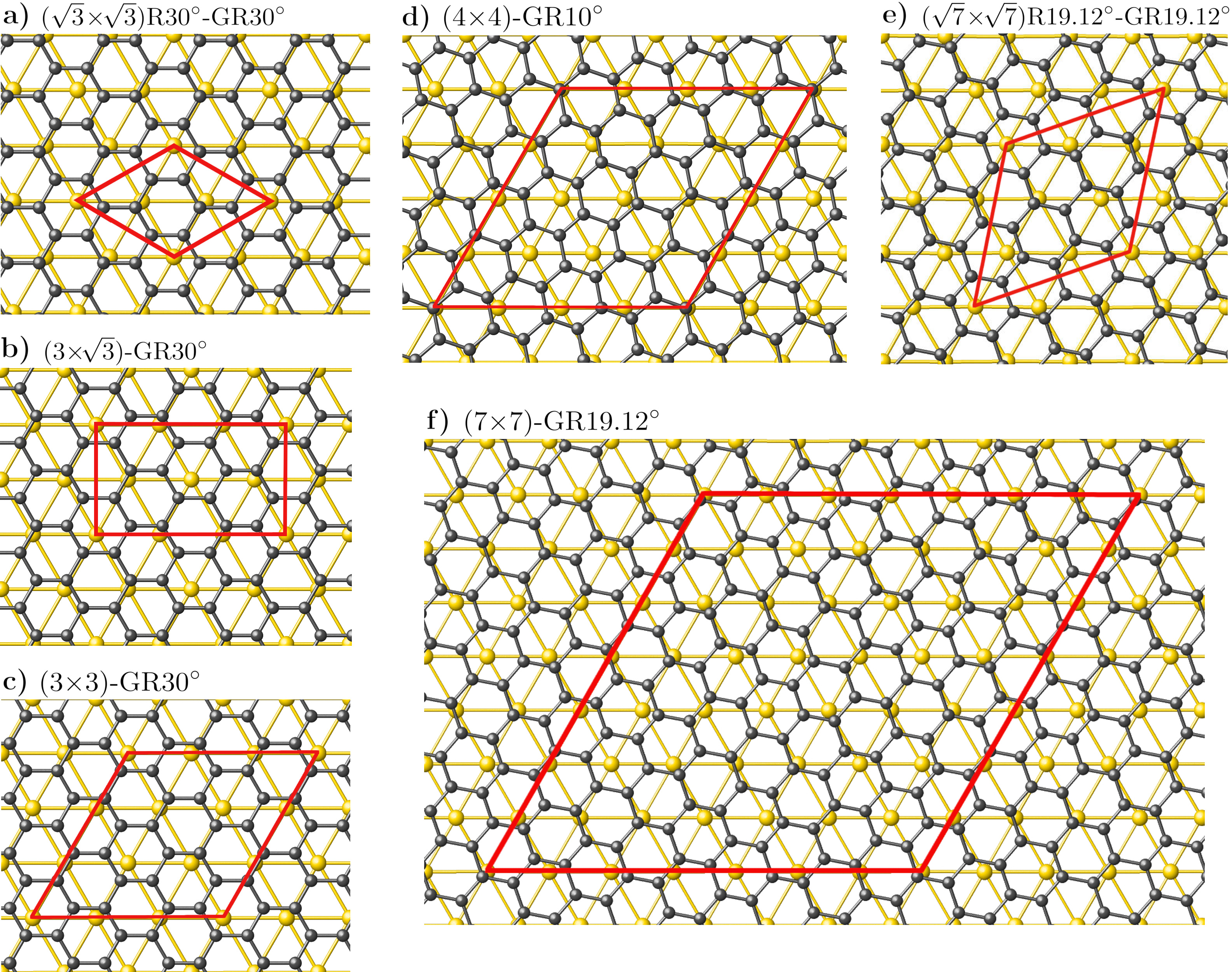}
\caption{(a) DFT-optimized common surface structures of graphene on 
gold (111) using the SCAN+rVV10 XC functional. The periodic supercell 
geometries were indicated by red (solid) lines. The labeling of common 
superstructures follows that of Au(111) on the left, a dash in the middle, and 
the rotational angle of graphene on the right.   \label{fig2}}
\end{figure*}

\section{Results and Discussion}

\subsection{Geometry and Adsorption}

Common periodic structures of graphene/Au(111) were modeled by overlapping 
the lattice vectors of gold and graphene unit cells, and then rotating the  
graphene monolayer on the Au(111) surface. The in-plane lattice vectors and 
initial configuration of both structures are shown in Fig.~\ref{fig1}. In this 
position, a common superstructure can be obtained which corresponds to 
graphene/Au(111)-($9\times 9$)/($8\times 8$) supercell. Due to its large 
size we did not include this geometry in this study. When the graphene 
monolayer is rotated 30$^{\circ}$ clockwise (or even counterclockwise) 
with respect to the in-plane lattice vectors of gold, it 
becomes commensurate with the ($3\times 3$) gold surface supercell
(in Fig.~\ref{fig2}c). We refer this geometry as ($3\times 3$)-GR30$^\circ$ 
indicating that graphene has a rotational angle of 30$^\circ$ on Au(111) with a 
common periodicity of ($3\times 3$) cell with respect to the gold surface. 
LEED experiments confirmed the existence of this R30 alignment 
structure.\cite{Wofford} At this rotational angle, other common 
superstructures of ($3\times\!\sqrt{3}$)-GR30$^\circ$ and 
($\sqrt{3}\times\!\sqrt{3}$)R30$^\circ$-GR30$^\circ$ can be identified as given 
in Fig.~\ref{fig2}a and Fig.~\ref{fig2}b with slightly different mismatch  
values. In particular,  ($\sqrt{3}\times\!\sqrt{3}$)R30$^\circ$-GR30$^\circ$ 
gold surface cell (Fig.~\ref{fig2}a) matches with ($2\times2$) supercell of 
graphene. The strain ratios for these three geometries are all about $-0.4$\% 
with the PBE and $-2$\% with the SCAN+rVV10 functionals as shown in 
Table~\ref{table2} where negative strain refers to shrinking.

%lattice mismatch (http://www.iue.tuwien.ac.at/phd/dhar/node12.html)
\begin{table*}[htb]
\caption{\label{table2} Minimum energy heights of graphene  
on Au(111) surface and the ratios of strain due to mismatch of graphene 
monolayer with Au(111) surface cell. Negative strain rates represent 
shrinking.}
\begin{ruledtabular}
\begin{tabular}{@{\hspace{2mm}}lcc@{\hspace{5mm}}cc@{\hspace{2mm}}}
\textbf{Structure} & \multicolumn{2}{c}{\textbf{PBE}} 
& \multicolumn{2}{c}{\textbf{SCAN+rVV10}} \\
&\textbf{h (\AA)} &\textbf{strain (\%)} & \textbf{h (\AA)}
&\textbf{strain (\%)}\\   
\hline \\[-4mm]
($\sqrt{3}\!\times\!\!\sqrt{3}$)R30$^\circ$-GR30$^\circ$ & 4.260 & $-$0.411 & 3.476 
& $-$2.019 \\
(3$\times\!\sqrt{3}$)-GR30$^\circ$ & 4.266 & $-$0.472 & 3.480 & $-$2.024 \\[1mm]
(3$\times$3)-GR30$^\circ$ & 4.258 & $-$0.404 & 3.480 & $-$1.989\\[1mm]
(4$\times$4)-GR10$^\circ$ & 4.238 & ~~0.377 & 3.459 & $-$1.244\\[1mm]
($\sqrt{7}\!\times\!\!\sqrt{7}$)R19.12$^\circ$-GR19.12$^\circ$ & 4.223 & ~~1.409 & 
3.448 & $-$0.228 \\[1mm]
(7$\times$7)-GR19.12$^\circ$ & 4.218 & ~~1.116 & 3.443 & $-$0.190\\
\end{tabular}
\end{ruledtabular}
\end{table*}

Since the angle between the in-plane lattice vectors (or rows of gold atoms) of 
gold is 60$^\circ$ giving a hexagonal arrangement, rotation of graphene monolayer 
by an angle of 10$^\circ$ or 50$^\circ$ results in the same geometry. Therefore, 
graphene/Au(111) ($4\times 4$)  superstructure shown in Fig~\ref{fig2}d can either 
be called as ($4\times 4$)-GR10$^\circ$ or equivalently as ($4\times 4$)-GR50$^\circ$ 

A larger, common geometry was identified at a rotational angle of 19.12$^\circ$ 
with a periodicity of ($7\times 7$) on the gold surface as depicted in 
Fig.~\ref{fig2}f.  At this rotational angle, another graphene/Au(111) superstructure 
can be identified with a slightly different mismatch value as 
($\sqrt{7}\times\!\sqrt{7}$)R19.12$^\circ$-GR19.12$^\circ$ (in Fig.~\ref{fig2}e). 
Consequently, a total of six different common supercells were determined with
low mismatch allowing probable Moir\'{e} patterns.

\begin{table*}[htb]
\caption{\label{table3} Average adsorption energy per C atom 
(E$_{\mathrm{ads}}$) , the work function of graphene, Au(111) and graphene 
coated gold surface ($\Phi$), the shift of the Dirac point of graphene with respect to 
the Fermi energy of the Au(111)/G system ($\Delta$E$_F$), the amount of charge 
(per C atom) displaced from graphene to gold ($\Delta$Q), calculated using the PBE 
and SCAN+rVV10 DFT functionals. Labeling of common structures of graphene and 
gold was made with respect to the surface unit cell of Au(111).}
\begin{ruledtabular}
\begin{tabular}{@{\hspace{5mm}}lccccccc@{\hspace{5mm}}}
\textbf{Structure} & \multicolumn{2}{c}{\textbf{PBE}} 
&\multicolumn{4}{c}{\textbf{SCAN+rVV10}}\\
&\hspace{3mm}\textbf{E$_{\textrm{ads}}$ (meV)} &\textbf{\hspace{2mm} $\Phi$ 
(eV)} &\hspace{3mm}\textbf{E$_{\textrm{ads}}$ (meV)} 
&\textbf{\hspace{2mm}$\Delta$E$_{F}$ (eV)}
&\textbf{\hspace{2mm}$\Phi$ (eV)} & \textbf{$\Delta$Q} \\[1mm]
\hline \\[-4mm]
($\sqrt{3}\!\times\!\!\sqrt{3}$)R30$^{\circ}$-GR30$^{\circ}$ & 
$-$1.64 & 4.85 & 
$-$62.66 & +0.18 & 5.17 & 0.00604e \\
(3$\times\sqrt{3}$)-GR30$^{\circ}$ & $-$2.51 & 4.84 & $-$64.74 & 
+0.17 & 5.16 & 0.00660e\\
(3$\times$3)-GR30$^{\circ}$ & $-$2.47 & 4.86 & $-$64.63 & +0.16 & 5.15 & 
0.00623e\\
(4$\times$4)-GR10$^{\circ}$ & $-$2.54 & 4.85 & $-$65.00 & +0.26 & 5.16 & 
0.00522e\\
($\sqrt{7}\!\times\!\!\sqrt{7}$)R19.12$^\circ$-GR19.12$^\circ$ & $-$6.02 &  4.86
& $-$70.13 & +0.34 & 5.12 & 0.00710e \\
(7$\times$7)-GR19.12$^{\circ}$ & $-$6.08 & 4.86 & $-$72.18 & +0.38 & 5.12 
& 0.00759e\\
Graphene & -- & 4.48 & -- & -- & 4.58 & -- \\
Au(111) & -- & 5.12 & -- & -- & 5.45 & -- \\
\end{tabular}
\end{ruledtabular}
\end{table*}

\begin{figure}[htb]
\includegraphics[width=7cm]{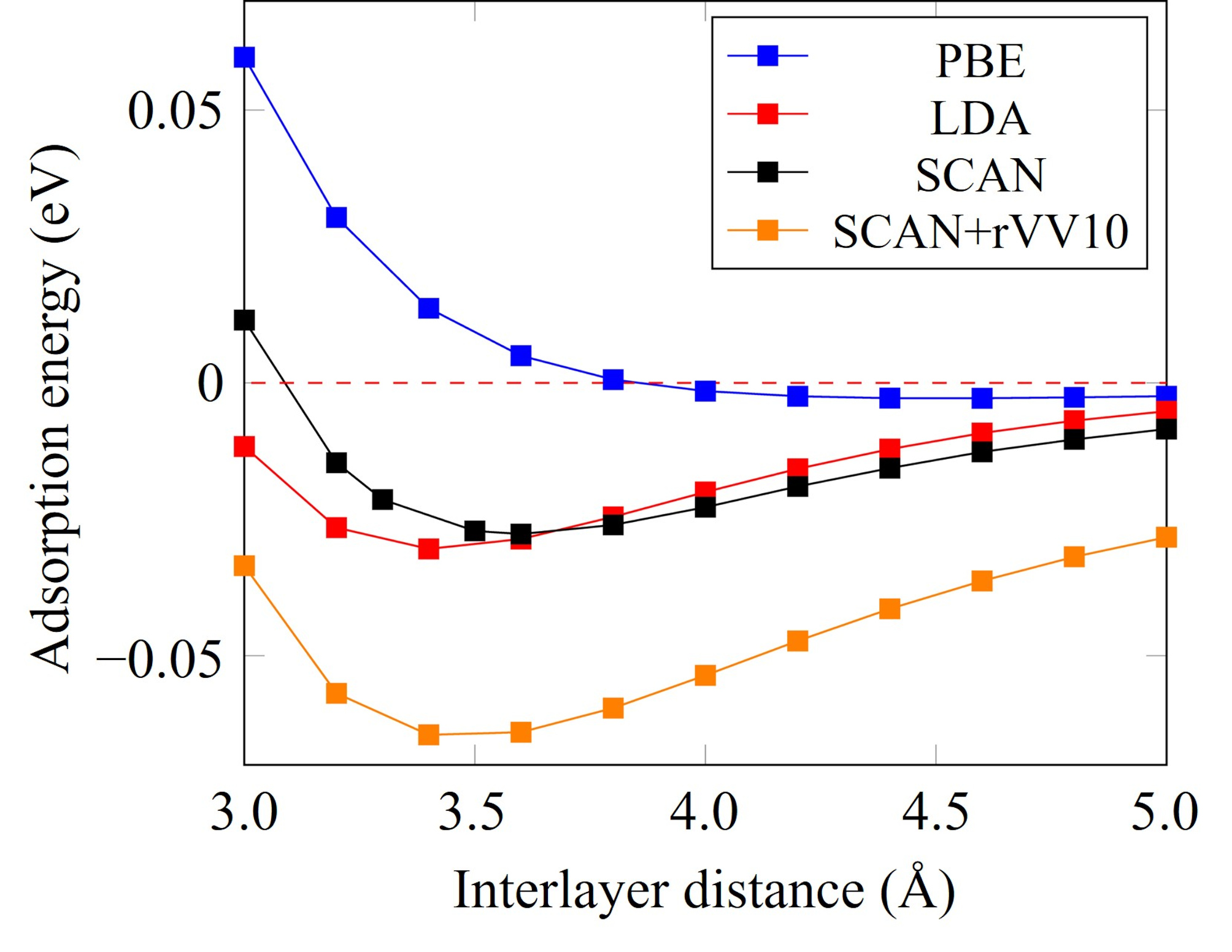}
\caption{Comparison of the adsorption energy (per C atom) 
profiles, calculated with different XC functionals, as a function of the 
interlayer distance between the graphene layer and the gold (111)-(4$\times$4) 
surface. \label{fig3}}
\end{figure}

Equilibrium distances and strain ratios of graphene monolayer on Au(111) 
surface were calculated using the PBE and the SCAN+rVV10 functionals, 
and presented in Table~\ref{table2}.  The height of graphene monolayer from 
the topmost gold plane represents an average value over the separation  
distances of carbon atoms from the gold surface layer. It is ranging between 
4.266 {\AA} and 4.218 {\AA} when calculated using the standard GGA XC 
functional. The SCAN+rVV10 functional not only includes vdW corrections
but also improves the description of many-body exchange-correlation effects.
Therefore, the height values become significantly smaller relative to those 
predicted by DFT-PBE calculations, ranging between 3.480 {\AA} and 
3.443 {\AA}. Previous vdW-DFT studies reported closely similar values using 
different vdW functionals.\cite{Khomyakov,Hamada,JZheng,Tesch,GZhao} We 
also note that some theoretical studies labeled the physisorption geometries 
with respect to graphene unit cell. For instance, 
Hamada~\textit{et al.}\cite{Hamada}, and 
Khomyakov~\textit{et al.}\cite{Khomyakov} labeled graphene/Au(111) adsorption 
geometries using ($1\times 1$) and ($2\times 2$) graphene supercells. 

Graphene is known to exhibit physisorption on the gold surface. 
\cite{Wintterlin,Tesch} In order to show the role of the vdW effects on the 
estimated values, we obtained the adsorption energy (per C atom) profiles as a 
function of the graphene/Au(111) interlayer distance using different XC 
functionals as shown in Fig.~\ref{fig3}. In these calculations, 
the distance of graphene monolayer from the gold surface was changed stepwise 
and no relaxation was performed. The PBE functional severely underestimates the 
adsorption energy. The standard GGA functionals are not suitable for  
chemisorption on metal systems either. For instance, even though the 
interaction between graphene and Ni(111) show a chemisorption behavior, PBE 
results suggest physisorption.\cite{Fuentes} The SCAN and LDA functionals give 
a relatively better but not sufficiently strong interaction between graphene 
and gold substrate. When the SCAN functional is supplemented with the rVV10  
vdW corrections, a significant improvement can be obtained on both the 
adsorption energy and the equilibrium distance. 

In addition, the adsorption energy profile for the PBE functional gets almost 
insensitive to interlayer distance after $\sim$4.2 {\AA}. Therefore, the 
PBE-predicted heights do not show a correlation with the strain ratios of 
graphene on gold. On the other hand, vdW-DFT results indicate a strong 
correlation between the height and the strain ratios as seen in 
Table.~\ref{table2}. The factors mentioned above necessitates the inclusion of 
dispersive forces in the theoretical calculations involving weakly interacting 
layered materials like graphene/Au(111).

\begin{figure*}[htb]
\centering
\includegraphics[width=15cm]{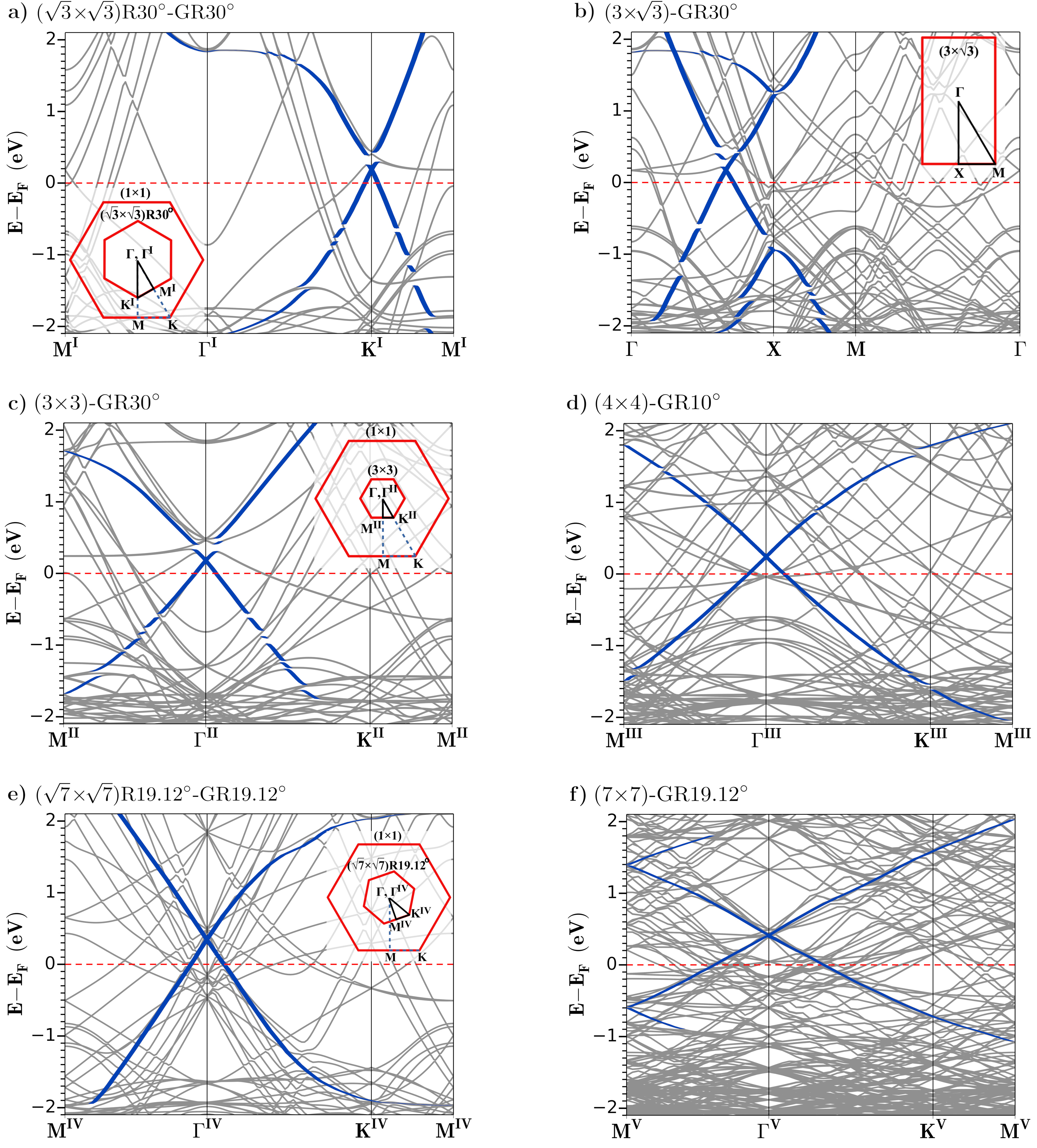}
\caption{Electronic band structures of graphene monolayer on Au(111) with
identified common superstructures. The corresponding two-dimensional BZ 
and high-symmetry  points are depicted in the insets. The contributions from 
graphene $p_z$ orbitals are highlighted with blue color (thick dark bands).
The energy values are given relative to the Fermi level (dashed lines) of 
each adsorption system.
\label{fig4}}
\end{figure*}

Calculated adsorption energies per carbon atom are given in 
Table~\ref{table3}. The PBE functional tends to underestimate the binding to a 
few meVs due to lack of long-range dispersive interactions as expected. The 
adsorption energies of a single layer graphene on the gold surface given by 
Zheng \textit{et al.}\cite{JZheng} for the PBE functional are in excellent 
agreement with our results in Table~\ref{table3}. On the other hand, vdW 
corrections bring a significant improvement over the energy related estimations.
The interaction of graphene on the metal is relatively weak with 
E$_{\mathrm{ads}}$=$-$62.66 meV at the
($\sqrt{3}\times\sqrt{3})$)R30$^\circ$-GR30$^\circ$ superstrucutre and gets 
stronger with the increasing supercell size up to E$_{\mathrm{ads}}$=$-$72.18 
meV at the (7$\times$7)-GR19.12$^{\circ}$ superstructure. The latter adsorption 
energy is in good agreement with a previous result obtained using DFT-D3 
method.\cite{Tesch} Depending on the choice of the vdW functional scheme, the 
corresponding estimations may slightly vary. Our adsorption energy values 
calculated using the SCAN+rVV10 functional gives a reasonable agreement 
with previous theoretical estimations.\cite{Andersen,JZheng}
The differences in the calculated adsorption energies between 
computational studies regarding the graphene/Au(111) system can be attributed 
to the use of different supercell structures and the vdW method. Another 
important factor appears to be the lack of rotational angular orientation of 
graphene on the gold surface in the previous studies. For instance, 
Tesch~\textit{et al.} considered the same alignment of graphene on both 
($2\times 2$) and ($7\times 7$) graphene supercells which correspond to 
($\sqrt{3}\times\!\sqrt{3}$)R30$^\circ$ and ($6\times\!6$) gold surface 
supercells, respectively. In our optimization calculations graphene gets bulged 
on ($6\times 6$) gold surface therefore we did not include this case.

As a result, the graphene/metal interlayer distance is sensitive to the rotational 
angle/alignment of graphene monolayer on the metal surface. Moreover, adsorption 
energies also show an almost linear correlation with the interlayer distances 
computed with the SCAN-rVV10 functional.

\subsection{Electronic Structures}

The electronic band structures of corresponding common graphene/Au(111)
systems were obtained using the SCAN+rVV10 functional as shown in
Fig.~\ref{fig4}. The Brillouin zone shapes, high symmetry points and the 
$k$-paths are shown in the insets. The size and the rotational alignment of 
the corresponding BZ are also depicted relative to the hexagonal BZ of 
the ($1\times 1$) surface unit cell of gold.  

It must be noted that the $k$-points ($\Gamma$, K , M, and X) are given
with respect to the BZ of gold supercell. There is also inward folding of the 
BZ when the real space periodicity increases. Therefore, the conicals not 
always appear at the K point of graphene unitcell. In addition, the BZ shapes 
of hexagonal ($3\times 3$), ($4\times 4$), and ($7\times 7$) supercells are 
similar. They only get smaller in size with increasing supercell size in the 
real space.

The weak interfacial interactions represented by vdW corrections showed 
that the Dirac conicals are preserved. The linear dispersion of graphene 
bands within $\pm$1 eV vicinity of the Dirac points of conicals can still be seen 
in all of the graphene adsorption orientations considered in this work. Moreover, 
graphene-metal interaction is weak such that it does not cause a band gap opening 
for the graphene monolayer. Then, the Dirac conicals differ in the position of the 
Fermi-level relative to the Dirac points. For a free-standing graphene,
the Fermi-energy must be at the Dirac point. When graphene gets closer to the 
Au(111) surface a Fermi-level shift is observed. The downward shift of Fermi 
energy with respect to the Dirac point corresponds to a $p$-type doping of 
graphene monolayer. 

The Fermi energy shift (with respect to the Dirac point) was found as +0.18 eV 
for $\sqrt3\times\!\sqrt3$)R30$^\circ$-GR30$^\circ$ using the SCAN+rVV10 
functional. This result is in good agreement with a previous value of +0.19 eV 
for the same structure.\cite{Khomyakov}  Vanin \textit{et al.} estimated a shift of 
+0.21 by using the vdW-DF method. Tesch~\textit{et al.}\cite{Tesch} synthesized 
graphene nanoflakes on Au(111) and reported a $p$-type doping with a Fermi 
energy shift of 0.24 $\pm$ 0.07 eV.\cite{Tesch} Similarly, Zheng \textit{et al.}
reported a Fermi-level shift of 0.230 eV for the bilayer graphene which is also
reasonably close to the calculated values presented in this work. 
Our results for the R30$^\circ$ alignment structures, ($3\times 3$)-GR30$^\circ$ 
and ($3\times\!\sqrt{3}$)-GR30$^\circ$, the shift values are similar because they all
reflect essentially the same surface characteristics. 

S\l{}awi\'{n}ska~\textit{et al.}\cite{Slawinska2} used scanning tunneling spectroscopy 
(STS) to show that Fermi-level shift ranges between 0.25 and 0.55 eV on different 
$p$-type doping domains of graphene on the (111) surface of gold. Our results for
the graphene/Au(111) physisorption systems, indicate that the shift changes with
the rotational alignment of graphene on the metal surface. The largest shift is 
calculated for the R19.12$^\circ$ alignment. The equilibrium distances of the 
graphene/Au(111) structures are still large for a charge transfer from graphene 
to gold surface. Then, we calculated Bader charges on graphene and on metal 
slab before and after adsorption for each of the structures considered in this work. 
Calculated charge transfer values from the graphene monolayer to the 
gold substrate are very small and only slightly larger in the case of 
R19.12$^\circ$ structures. Hence, a charge transfer model may not fully explain 
the Fermi-level shift mechanism. Our results rather suggest a charge 
redistribution induced by graphene-metal interlayer interactions. 

The change in the Fermi-level shift between different common graphene/Au(111) 
geometries can be attributed to the rotational angle. Because, different angles cause
different atomic coincidences of C atoms with the surface gold atoms. This also
leads to a change in the ratio of the number of C atoms to the number of surface 
Au atoms ($n_{C/Au}$) contained in the supercell. This ratio is $n_{C/Au}$=2.67, 
2.63, and 2.57 at the rotational angles of 30$^\circ$, 10$^\circ$, and 19.12$^\circ$, 
respectively. 

Another factor affecting the Fermi-level shift can be the work function 
of graphene/Au(111).  Theoretical studies usually explain the shift as the 
difference between the work functions of metal surface and graphene monolayer.
Therefore, we calculated the work functions of common superstructures. 
The real space potential energy profiles of the common hetero-layered structures 
were obtained along the [111] direction. Then the work function is 
the difference between the Fermi energy of the combined system and 
the vacuum level. Experimental work function of graphene and the gold surface 
is $\sim$4.62 eV.\cite{Song,Yuan} and 5.31 eV, respectively. Our 
SCAN+rVV10 functional estimation of 5.45 eV for the clean gold surface is 
in good agreement with recent theoretical values.\cite{Patra} The PBE 
functional, on the other hand, gives a value of 5.12 eV for Au(111). Graphene 
monolayer is estimated to have a work function of 4.58 eV and 4.48 eV with and 
without the vdW-corrections, respectively. Calculated work 
function values of graphene coated gold surface show a significant reduction 
relative to the that of the clean gold surface. Furthermore, our results does 
not show a meaningful trend to relate with the Fermi-energy shifts. On the 
other hand, the $\Phi$ values for different structures all exhibit similar 
characteristics. Therefore, the work functions of graphene/Au(111) systems 
appear to be independent of graphene monolayer orientation on the gold surface. 
Interestingly, Song~\textit{et al.} recently reported that graphene in contact 
with Au gets an intermediate work function value and still dominantly show 
features of graphene.\cite{Song} The theoretically estimated work function 
values of the graphene/Au(111) superstructures assume an intermediate value and 
not pinned to the work function of the metal. 

\section{Conclusions}

We have identified a number of small-sized common graphene/Au(111) 
superstructures by matching their corresponding supercells. Structural 
optimizations and electronic properties were obtained by performing DFT 
calculations. The weak interaction at graphene-gold interfaces necessitates 
the inclusion of vdW corrections. The standard GGA functionals tend to 
severely underestimate adsorption energies. The common geometries show 
different equilibrium distances between the graphene monolayer and the gold 
surface depending on the rotational orientation between them. 

Graphene adsorption on gold does not destroy the Dirac conicals which still
show linear dispersion within $\pm$1 eV vicinity of the Dirac points. 
A change in the Fermi energy indicate $p$-type doping of graphene when 
adsorbed on the metal. Most importantly, we showed that the Fermi-level shifts
depend on the rotational alignment of the graphene on the gold surface.

Consequently, the results showed that the adsorption and electronic properties 
of graphene/Au(111) system are sensitive to the matching of graphene 
monolayer to the geometry of the gold surface at different rotational angles.
This would allow a tunability of characteristics of potential applications based 
on  graphene/metal layered materials. The common superstructures identified in 
this study are useful and serve as building blocks to construct and design 
novel hetero-layered materials involving graphene and Au(111).

\section*{Acknowledgements}
This study was supported by Tubitak, The Scientific and Technological Research 
Council of Turkey, under Grant No. 116F174. The calculations were performed 
on the High Performance and Grid Computing Center (TRUBA). The authors also 
acknowledge financial support from Bal{\i}kseir University under Grant No. BAP 
2018/039.

%\nocite{*}
\bibliography{AuG}

\end{document}